\documentclass[12pt]{article}  
\newcommand{\comma}{\;\; ,}
\newcommand{\period}{\;\; .}
\newcommand{\semicolon}{\;\; ;}
\newcommand{\eq}{\; = \;}
\newcommand{\sep}{\;\; , \;\;}
\newcommand{\be}{\begin{equation}}
\newcommand{\bd}{\begin{displaymath}}
\newcommand{\ee}{\end{equation}}
\newcommand{\ed}{\end{displaymath}}
\newcommand{\ba}{\begin{eqnarray}}
\newcommand{\ea}{\end{eqnarray}}

\newcommand{\minus}{\! - \!}

\newcommand{\plus}{\! + \!}
\newcommand{\half} {\textstyle{\frac{1}{2}}}

\newcommand{\rii}{{\rm i}}
\newcommand{\oQ}{\tilde{Q}}
\newcommand{\rt}{\raisebox{.4ex}{$\scriptstyle{rt}$}}
\newcommand{\ts}{\raisebox{.4ex}{$\scriptstyle{ts}$}}
\newcommand{\rs}{\raisebox{.4ex}{$\scriptstyle{rs}$}}
\newcommand{\sr}{\raisebox{.4ex}{$\scriptstyle{sr}$}}
\newcommand{\tr}{\raisebox{.4ex}{$\scriptstyle{tr}$}}

\title{The six and eight-vertex models revisited }

\author{ R.J. Baxter
\footnote{Mathematical
Sciences Institute, The Australian National University,
 Canberra, A.C.T. 0200, Australia}
\footnote{This 
work supported in part by 
the Australian Research Council.} }

\date{4 September 2003}

\begin{document}

%\magnification = \magstep1
%\magnification = 1000

\maketitle

\abstract{Elliott Lieb's ice-type models opened up the whole
field of solvable models in statistical mechanics. Here we
discuss the ``commuting transfer matrix'' $T, Q$  equations 
for these models, writing them in a more explicit and transparent 
notation that we believe offers new insights. The approach manifests 
the relationship between the  six-vertex and chiral Potts
models, and between the eight-vertex and Kashiwara-Miwa models.

}

\section{INTRODUCTION}
%%3456789012345678901234567890123456789012345678901234567890123456789012
Elliott Lieb used the Bethe ansatz to solve the two-dimensional ice, 
F and KDP models in 1967 - 
all typical cases of the general ``six-vertex'' model whose solution 
was then given by Sutherland.\cite{Lieb67a} - \cite{Sutherland67} This 
work, together with Onsager's famous 
solution \cite{Onsager44} of the square-lattice Ising model in 1944, 
formed the basis on 
which the field of two-dimensional models rapidly grew. It led the 
author to develop the ``commuting transfer matrix'' of tackling such 
problems, notably the eight-vertex  and hard-hexagon models.

This method begins with two steps. The first is to treat the problem in 
sufficient generality that it contains an arbitrary parameter (a 
complex  number) $v$ such that the transfer matrix $T$ is a 
non-trivial function $T(v)$ of $v$, and
\be \label{TTcommn}
 T(u) T(v) = T(v) T(u) \ee
for all values of $u$ and $v$. Thus $T(u)$ and $T(v)$ commute.
The variable $v$ is known as the ``spectral parameter'': for the six 
and eight-vertex models (and many other planar models) it is the 
difference of two ``rapidity'' variables associated with the horizontal 
and vertical directions of the lattice.

The second step is to construct (if only implicitly) another matrix
function $Q(u)$ which also commutes with $T(v)$, i.e.
\be \label{QTcommn}
Q(u) T(v) = T(v) Q(u) \comma \ee
and satisfies a matrix functional relation that is linear in $T$ and 
$Q$, and homogeneous in $Q$. For the six-vertex model this has the 
structure
\be \label{fnlreln}
T(v) Q(v) \eq  \phi (v-\lambda) Q(v+2 \lambda') + 
\phi(v+\lambda) Q(v-2 \lambda') 
\comma \ee
$\phi(v)$ being a known scalar function and $\lambda'$ a 
``crossing parameter''.

A further step is to show that  $Q(u)$ and  $Q(v)$ commute:
\be \label{QQcommna}
Q(u) Q(v) = Q(v) Q(u) \comma \ee
for all $u, v$. 

The final step  is to go to a 
representation in which $T(v), Q(v)$ are diagonal matrices, for all 
$v$. Then (\ref{fnlreln}) becomes a scalar functional relation for each 
eigenvalue, which can be solved.

%%3456789012345678901234567890123456789012345678901234567890123456789012

For the six and eight-vertex models, all these steps have been carried 
out by the author \cite{Baxter71,Baxter72,Baxter73} and written up in 
chapters 9 and 10 of \cite{book82}. However, the construction of $Q(v)$ 
given therein is implicit rather than explicit. 

Further, in the early papers \cite{Baxter71,Baxter72,Baxter73} the 
author was concerned to focus on cases  where there was no reason to 
believe the matrix $Q(v)$ was in general 
singular. Obviously if it is zero the equations  (\ref{QTcommn}) - 
(\ref{QQcommna}) contain no information on $T(v)$. Only if it is 
non-singular (i.e. has non-zero determinant) for general values of 
$v$ can one expect to obtain 
all the eigenvalues of $T(v)$ from (\ref{fnlreln}). For this reason 
the author initially focussed on the  ``roots of unity'' cases, where 
one can write $Q(v)$ more explicitly. Later he was able to remove this 
restriction, and the working in \cite{book82} is a quite general 
solution of the zero-field six and eight-vertex models.

%%3456789012345678901234567890123456789012345678901234567890123456789012

Here we re--present this working for the six-vertex model, giving a more 
explicit expression for the matrix $Q(v)$.\footnote{Basically our 
results are more explicit versions of equations (9.8.15) - (9.8.37) of 
\cite{book82}.} The relations  
(\ref{TTcommn}) -- (\ref{QQcommna}) are derived as the set of equations
(\ref{final}), where the skew parameter $r$ is exhibited and 
(for $L$ even) the other skew or field parameter $s$ can be 
given the value 0. 

The formulae (\ref{WQ}), 
(\ref{deftQ}) for the transfer matrix $Q_R(v)$  are interesting 
in that it is almost that of a trivial one-dimensional 
model: the only thing that  stops this happening is  a factor 
$x^{d (a-b)}$  in the Boltzmann weight function (\ref{WQ}). 
A very similar property occurs in the three-dimensional Zamolodchikov 
model and its extension.\cite{RJB1986,BazBax93}

In 1990 Bazhanov and Stroganov \cite{BazStrog90} showed how the
 recently-discovered
solvable chiral Potts model could be obtained from the six-vertex model
via the matrix $Q(v)$. They used the  general $Q$ matrix discussed in
section 6 (also including the fields we mention at the end of our 
conclusion).\footnote{It is the $\tau_2(t_q)$ matrix of \cite{BBP90}, 
except that it is rotated through  $90^{\circ}$. Bazhanov and Stroganov rotated the 
lattice and started with the column-to-column transfer matrix of the 
six-vertex model. This leads to the usual row-to-row transfer matrix 
of the chiral Potts model. Here we shall not be extending the 
calculation that far, so revert to using the row-to-row 
transfer for the six-vertex 
model.} In this approach the Boltzmann weight of the chiral Potts model 
first appears as an auxiliary function that enters the derivation of 
(\ref{QQcommna}). Here we shall observe such a property.

%%3456789012345678901234567890123456789012345678901234567890123456789012

We emphasize that our working is for {\em all} values of the crossing 
parameter $\lambda$ or $\lambda'$. This means that the spins in our 
spin formulations of the six-vertex and $Q$ models are free 
to take all integer values from $-\infty$ to $ + \infty$. We use this 
convenient ``spin language '' to derive the relations, but then in 
section  5 we indicate how one can transform back to the arrow 
language of Lieb so as to ensure that the row-to-row transfer matrices 
are  finite-dimensional.

Only in the conclusion do we address the ``roots of unity'' cases, when
$\lambda' = {\rm i} m  \, \pi/N$,   $x^{4N} = 1$, ($m,N$ integers)
 and the spin on a given site takes just $N$ states.
These cases are of course of interest. Fabricius and McCoy  
\cite{DegFabMcCoy01} -  \cite{FabMcCoy01c} have studied these cases 
and have emphasized that the eigenvalues of $T(v)$ are then  
degenerate.\footnote{Fabricius and McCoy use the word ``incomplete''
in a way quite different from its usual sense in mathematical physics
\cite{Baxter02,Dirac,Weyl}.}
This provides a motivation for establishing 
the $Q, Q$  commutation relations directly: if the eigenvalues 
of $T^r(v)$ are non-degenerate, the $Q, Q$ relations are implied by 
those for $T,T$ and $T,Q $. Further, it is this step that first 
introduces the weight function of the chiral Potts model. Only for 
the $N$-state root-of-unity case does this model seem to be properly
defined.

Finally, we briefly indicate how the same approach applied to the 
eight-vertex model leads quite directly to the Kashiwara-Miwa 
model.\cite{KashMiwa86} As Hasegawa and 
Yamada have shown\cite{HasYam90}, the Kashiwara-Miwa model is a 
``descendant'' of the zero-field eight-vertex model in the same way
that the chiral Potts model is a ``descendant'' of the six-vertex 
model in external fields. 

This answers a question that has long 
puzzled the author: can one obtain a generalization of chiral Potts 
by using the Bazhanov-Stroganov method, but starting with the 
eight-vertex model instead of the six-vertex? The answer appears to 
be no. The model from which Bazhanov and Stroganov start is the 
six-vertex model in a field. Hasegawa and 
Yamada start from the zero-field eight-vertex model.  There is 
no known solvable model that continuously generalizes and includes 
both these vertex models, so there is no reason to suppose that 
there exists a model which similarly generalizes and includes the 
chiral Potts and Kashiwara-Miwa models.

It is true that the two vertex models intersect in the zero-field 
six-vertex model. The ``descendant'' of this is the 
Fateev-Zamodochikov model\cite{FatZam82}, which is  indeed the intersection of the
chiral Potts and Kashiwara-Miwa models.

The relationships between these  six models are shown in Figure \ref{models}.

\section{TRANSFER MATRICES AND THE STAR-TRIANGLE RELATION}

Here we make some general observations to which we shall refer in
subsequent sections.

We shall consider various  ``interactions-round-a-face'' (IRF) models 
on the square lattice $\cal L$ of $\cal N$ sites. 
In each model,  each site of the lattice carries a ``spin''. Let $a, 
b, c, d$ be the four spins  round a face, arranged as in 
Figure \ref{fourspins}, and let  $W(a, b , c, d )$ be the 
Boltzmann weight of this spin configuration. Then the partition 
function  is 
\be Z \eq \sum \prod W(a, b , c, d ) \comma \ee
the product being over all faces and the sum over all allowed values of 
all the $\cal N$ spins.\footnote{Some of the spins may need to be 
fixed to avoid this sum being infinite. For the six-vertex model just 
one spin on the lattice needs to be fixed - say the one at the 
bottom-right corner. For the ``$Q$'' model we shall discuss we need 
to fix one spin in every row - say the left-most spin.} There are 
usually restrictions on the spin  values: here we shall require that 
on {\em some} (not necessarily all) edges
$(i,j)$ the two adjacent spins $\sigma_i, \sigma_j$ differ by unity:
\be \label{adjspins} 
\sigma_j =  \sigma_i \pm 1 \period \ee

In particular, we require that (\ref{adjspins}) be satisfied on all 
horizontal edges. Let the lattice have  $L$ columns, i.e. $L$ faces 
per row, and let the spins in any particular row be 
$\sigma_1, \sigma_2, \ldots ,
\sigma_{L+1}$. Then we impose skewed cyclic boundary conditions:
\be \label{defskewparm}
\sigma_{L+1} = \sigma_1 + r \comma \ee
where $r$ is the ``skew parameter'' - an integer with value between 
$-L$ and $L$ such that $L-r$ is even. It may vary from row to row.

We shall also use the spin differences:
\be \label{defalpha}
\alpha_i = \sigma_{i+1} - \sigma_i \period \ee
we see that each $\alpha_i$ takes the values $+1$ or $-1$, and
\be \label{sumalphas}
\alpha_1 + \cdots + \alpha_L \eq r \period \ee

Let $\sigma = \{ \sigma_1, \ldots , \sigma_{L+1} \} $ be the set of 
spins in one row, and let
$\sigma' = $ $\{ \sigma'_1, \ldots , \sigma'_{L+1} \} $ be the 
spins in the row above, as in Figure \ref{row}. Then the 
row-to-row {\em transfer matrix} $T$ is the matrix with elements
\be \label{defT}
T_{\sigma, \sigma'} \eq W(\sigma_1,\sigma_2,\sigma'_2,\sigma'_1 ) 
\, W(\sigma_2,\sigma_3,\sigma'_3,\sigma'_2 )  \ldots 
W(\sigma_L,\sigma_{L+1},\sigma'_{L+1},\sigma'_L )  \period \ee
If the lattice has $M$ rows, with cyclic top-to-bottom boundary
conditions, it follows that
\be
Z \eq {\rm Trace} \;  T^M \period \ee

Obviously $T$ depends on the skew parameter $r$.
If $r$ is the same for both rows, we may write $T$ as $T^r$.
If it is different, so that $\sigma_{L+1} = \sigma_1 + r$ and
$\sigma'_{L+1} = \sigma'_1 + s$, we may write $T$ as $T^{rs}$.

\subsection*{Commutation}

Now consider two models, with different weight functions $W_1$ and 
$W_2$. Let their transfer matrices be $T_1$, $T_2$ (for the moment we 
suppress the superfixes $r,s$). The product  $T_1 T_2$ has entries
\be \label{trace}
[T_1 T_2]_{a,b} \eq {\rm Trace} \; \; M_1 M_2 \cdots M_L \period \ee
Here $M_j$ is a a matrix which depends on four outer spins 
$a_j, a_{j+1} , b_j, b_{j+1}$ and has entries
\be \label{defM}
[M_j]_{c,d} = W_1(a_j, a_{j+1},d,c) W_2(c,d,b_{j+1},b_j ) \period \ee
The entries of $T_2 T_1$ are similar, but $M_j$ is replaced by $M'_j$,
in which $W_1$ and $W_2$ are interchanged. For $T_1$ to commute with 
$T_2$, i.e. for $T_1 T_2 = T_2 T_1$, we need there to exist invertible
matrices $P_1, \ldots, P_{L+1}$ such that $P_j$ depends on the outer 
spins $a_j, b_j$ (and no others), and
\be \label{condn}
 M'_j \eq P_j^{-1} M_j P_{j+1} \; \; \; , \; \; \; j = 1,\ldots, L
 \period \ee

Remembering that the matrix $P_j$
depends on $a_j, b_j$, write its entry $(c,d)$ as $W_3(c,a_j,d,b_j)$.
Then, rewriting (\ref{condn}) as   $M_j P_{j+1} = P_j M'_j$,
we obtain this condition explicitly as  
\bd
\sum_g  W_{1}(b,c,g,a )  W_{2}(a,g,e,f )  W_{3}(g,c,d,e )
\eq \ed
\be \label{startri}
\sum_g W_{3}(a,b,g,f ) W_{2}(b,c,d,g)  W_{1}(g,d,e,f)   \ee
for all allowed values of the external spins $a, b, c, d, e, f$. 

For interaction-round-a-face models, (\ref{startri})  is the 
``star-triangle'' relation. It is depicted graphically in 
Figure \ref{startrifig}.

For cyclic boundary conditions, $P_{L+1} = P_1$ and 
(\ref{startri}), together with the invertibility of
$P_j$, is sufficient to ensure that $T_1 T_2 = T_2 T_1$, i.e.
$T_1, T_2$ commute.

%%3456789012345678901234567890123456789012345678901234567890123456789012

However, for our more general skewed boundary conditions we do have a
problem. Including the skew parameters,  the commutation relation 
in general becomes
\be \label{gencommn}
T^{rt}_1 T^{ts}_2 = T^{rt'}_2 T^{t's}_1 \period \ee
Here $r$ is the skew parameter of the lowest row, $s$ of the uppermost,
and $t$, $t'$ of the intervening rows (the ones whose spins are summed 
over in the matrix multiplication). Note that $t$ is not 
necessarily the same as $t'$. If $P_1$ has entries  $W_3(c,a_1,d,b_1)$,
then $P_{L+1}$ has entries  $W_3(c+t,a_1+r,d+t',b_1+s)$.
If these entries are equal then (\ref{gencommn})
holds true.

In fact they  are not always equal: in section 4 we 
encounter a case where 
\be  \label{W3cyc}
W_3(c+t,a+r,d+t',b+s) \eq x^{s(c-a)} \; W_3(c,a,d,b) 
\comma \ee
$x$ being the fixed parameter defined in (\ref{defx}).
The effect of this is to insert an extra factor
$x^{s(c_1-a_1)}$ into the sum over $c_1, \ldots , c_L$ that is
the matrix product $T_1 T_2$. This is equivalent to replacing 
$T_1$ therein by $D^{-s} T_1 D^s$, where $D$ is a 
a diagonal matrix with entries
\be \label{defD}
D_{\sigma, \sigma'} \eq x^{\sigma_1} \, \delta(\sigma_1, \sigma'_1)
\cdots \delta(\sigma_{L+1}, \sigma'_{L+1}) \period \ee
(Note that here $\sigma$ and $\sigma'$ necessarily have the 
same value of the skew parameter $\sigma_{L+1} -\sigma_1$.)

Thus instead of (\ref{gencommn}) we obtain
\be
 \label{modcommn}
D^{-s} T^{rt}_1 D^s T^{ts}_2 \eq T^{rt'}_2 T^{t's}_1 \period \ee

%%3456789012345678901234567890123456789012345678901234567890123456789012

%%3456789012345678901234567890123456789012345678901234567890123456789012

\section{THE SIX-VERTEX MODEL}

As considered by Lieb and Sutherland, the six-vertex model is a model 
on the square lattice where one puts an arrow on every edge. At each 
site one is required to satisfy the ``ice rule'' that there be two 
arrows pointing in and two pointing out of the site. There 
are then six possible configurations of arrows at a site, as indicated 
in Figure \ref{sixv}.

For our present purposes, we wish to express this in 
interaction-round-a-face language. We do this by putting spins on the 
faces of the lattice so that as one goes from one face to a 
neighbouring  face, if the intervening arrow points to one's 
left (right), then the  spin value increases (decreases) by unity.

As one walks around a site, the ice rule ensures that there are two
arrows pointing to  the left, and two to the right, so one does indeed
return to the original spin value. 

It follows that a spin configuration is allowed iff every pair of 
adjacent spins (horizontal {\em and} vertical) differs by unity, 
i.e. satisfies (\ref{adjspins}).
If one of the spins is fixed (say the 
one at the lower-right-hand corner), then there is a one-to-one
correspondence between allowed arrow and allowed spin configurations 
on the lattice.

Finally, we go to the dual of the arrow lattice. The spins now live on
sites (rather than faces), and Figure \ref{sixv} becomes Figure
\ref{sixvertex}.

%%3456789012345678901234567890123456789012345678901234567890123456789012

The spin differences $\alpha_i$ mentioned above now describe the 
vertical arrows in a row: $\alpha_i = +1$  if the arrow between
spins $\sigma_i$ and $\sigma_{i+1}$ is pointing upwards, 
$\alpha_i = -1$ if it is pointing downwards. 
 We identify the horizontal arrow between 
$\sigma_1$ and $\sigma'_1$ with the  arrow between  
$\sigma_{L+1}$ and $\sigma'_{L+1}$ (i.e. in the arrow formulation 
we use the usual cyclic boundary conditions). Then the ice rule implies 
that if   (\ref{sumalphas}) is satisfied for one row, then it is
satisfied for all rows, with the {\em same} value of $r$. This is the
``conservation of up and down arrows'' 
property of the six-vertex model.\cite{Lieb67a} - \cite{Sutherland67} 

Let $w_1, \ldots ,w_6$ be the Boltzmann weights of the six 
configurations in Figure \ref{sixv} and Figure \ref{sixvertex}.
Here we consider only the ``zero-field'' six-vertex model, which is 
invariant under reversal of all arrows, so we take:
\be  w_1 = w_2 \sep w_3 = w_4 \sep w_5 = w_6 \period \ee

The overall normalization of $w_1, w_2, w_3$ plays a trivial role in
the calculations, so without loss of generality we can define two 
parameters $\lambda, v$ so that
\be w_2 = \sinh \half (\lambda-v) \sep  w_4 = \sinh \half (\lambda+v)
\sep w_6 = \sinh \lambda \ee
as in (9.2.3) of \cite{book82}. It follows that the Boltzmann weight 
function of the zero-field six vertex model is
\bd W(a,b,c,d) \eq W_{6v}(v|a,b,c,d) \eq \delta(a,c) \, 
{\rm e}^{(\lambda+v)(b+d-a-c)/4} \, 
\sinh\half (\lambda-v)   + \ed \be \label{defWsv} 
\delta(b,d) \, {\rm e}^{(\lambda-v)(a+c-b-d)/4} \, 
\sinh\half (\lambda+v) \period  \ee

We regard $\lambda$ (the ``crossing parameter'') as a given constant 
and  $v$ as a variable (the ``spectral parameter'').
We therefore write the six-vertex model transfer matrix as 
$T(v) = T^r(v)$ and define it as in  (\ref{defT}): 
\be \label{defT6v}
[T^r(v)]_{\sigma, \sigma'} \eq W_{6v}
(v|\sigma_1,\sigma_2,\sigma'_2,\sigma'_1 ) \ldots 
W_{6v} (v|\sigma_L,\sigma_{L+1},\sigma'_{L+1},\sigma'_L )  \period \ee
The lower and upper spin sets $\sigma, \sigma'$ have the same 
value of the skew parameter $r$.
 
The star-triangle relation (\ref{startri}) is satisfied if
\bd W_1 = W_{6v}(v) \sep W_2 = W_{6v}(v') \sep 
W_3 = W_{6v}(v' - v - \lambda) \ed
for all values of $v$ and $v'$. Also, taking $r ,s, s', t = r$,
we see that (\ref{W3cyc}) is satisfied by (\ref{defWsv}), with $x = 1$.

The six-vertex model transfer matrices therefore commute:
\be \label{svcommn}
T^r(v) \, T^r(v') \eq T^r(v') \, T^r(v) \period \ee

Multiplying $W_{6v}(a,b,c,d)$ by a field factor
$ \mu^{c-b} $  is equivalent to post-multiplying $T^r(v)$
by a diagonal matrix $S^z$ with entries
\be S^z_{\sigma, \sigma'} \eq \zeta(\sigma) \, \prod_{j=1}^{L+1}
\, \delta(\sigma_j, \sigma'_j ) \comma \ee
where \bd  \zeta(\sigma) \eq  \prod_{j=1}^L \mu^{\sigma_{j+1}-\sigma_j} \eq 
\mu^r \period \ed

This in turn corresponds to introducing an electric field acting 
on the upper vertical arrows, or equivalently on the upper pairs 
of horizontally adjacent spins. Clearly $S^z$ commutes with
$T^r(v)$:
\be S^z \, T^r(v) \eq T^r(v) \, S^z \period \ee

%%3456789012345678901234567890123456789012345678901234567890123456789012

\section{THE ``$Q$'' MODEL}

%%3456789012345678901234567890123456789012345678901234567890123456789012

The step taken by Bazhanov and Stroganov \cite{BazStrog90} was to
look for another solution of  (\ref{startri}), in which $W_3$ remains
unchanged, but $W_1$, $W_2$ are altered so that they are no longer
weight functions for the six-vertex model, but for a new 
``$Q$''-model. These functions can in fact be obtained 
(after changing the normalization) from
(9.8.15) - (9.8.23) of \cite{book82}. They are
\be \label{WQ}
W_Q(v|a,b,c,d) \eq \exp[{\lambda' d (a-b)/2}  - 
(\lambda+v)(b-a)(c-d)/4] \comma \ee 
where
\be \lambda' \eq \lambda - {\rm i} \pi  \comma \ee
 as in eqn. 9.3.7 of \cite{book82}.

Horizontally adjacent spins must still satisfy (\ref{adjspins}), so
$|b-a| = |c-d| = 1$, but now there are no restrictions on 
the differences  $d-a$ or $c-b$ of  vertically adjacent spins. Hence 
the skew parameters $r, s$ of the lower and upper rows in Figure 
\ref{row} can be different.

 We define the corresponding transfer matrix $Q_R^{\rs}(v)$ as 
in (\ref{defT}):
\be \label{deftQ}
[Q_R^{\rs}(v) ]_{\sigma, \sigma'} \eq 
W_Q(v|\sigma_1,\sigma_2, \sigma'_2,\sigma'_1) \cdots  
W_Q(v|\sigma_L,\sigma_{L+1}, \sigma'_{L+1} ,\sigma'_L) \period \ee

\subsection*{$T, Q$ commutation.}

In the star-triangle relation  (\ref{startri}) we now substitute 
\be \label{TQstartri}
W_1 = W_{6v}(v) \sep W_2 = W_Q(v') \sep W_3 = W_Q(v' - v - \lambda) 
\period \ee
To be consistent with the definitions of $W_{6v}$ and $W_Q$, we 
require that spins linked directly by edges in Figure 
\ref{startrifig} differ by one, {\em except} for the edges
$(a,f)$, $(g,e)$, $(c,d)$ on the lhs, and edges $(a,f)$, $(b,g)$, 
$(c,d)$ 
on the rhs. This means that the six external spins break up into
two sets: $(a,b,c)$ and $(d,e,f)$. One is free to independently
increment all the spins in either set by unity. We 
find that (\ref{startri}) is satisfied by (\ref{TQstartri}).

For the $T$ and $Q_R$ matrices to commute, we also need to 
check the 
the auxiliary condition (\ref{W3cyc}). Since $T$ necessarily
relates two rows of spins with the same skew parameter, we must have
$t = r, t'= s$ therein. We find that (\ref{W3cyc}) is satisfied, 
but $x$ therein is {\em not} unity: instead 
\be \label{defx}
x  = {\rm e}^{\lambda' /2} = -{\rm i} \; {\rm e}^{\lambda/2}
\period \ee

We therefore obtain the modified commutation relation 
(\ref{modcommn}), with $T_1$ replaced by $T^r(v)$ and $T_2$ by
$Q_R^{\rs}(v')$, i.e.
\be \label{TQcommn}
D^{-s} \, T^{r} \! (v) \, D^s \, Q_R^{\rs}(v') \eq 
Q_R^{\rs}(v') \, T^{s}(v) \period \ee
Here $v$ and $v'$ are arbitrary, $D$ is defined by (\ref{defD}) 
and (\ref{defx}). The notation is threatening to become confusing:
note that $r$ and $s$ here are merely superfixes, except that $D^s$ is
actually the diagonal matrix $D$ raised to the power $s$.
 
\subsection*{The $T, Q$ functional relation.}

Our working in section 2 depended on the matrices $P_j$ being
invertible. This will cease to be so if 
$W_3(a,b,c,d)$ factors into a function independent of $a$ times 
a function independent of $c$. For the derivation of (\ref{TQcommn})
this happens when $v' = v$.

This does not mean that the commutation relation fails - 
it follows by taking the limit $v' \rightarrow v$. It does mean
that there is then additional information in the relation 
(\ref{startri}).
The lhs then involves $d$ only via the simple factor $x^{e(g-c)}$, 
while the rhs involves $a$ only via $x^{f(a-b)}$.

Going back to equations (\ref{trace}) - (\ref{condn}), define
 vectors $\xi_1, \ldots \xi_L$ with elements
\be [\xi_j]_c \eq x^{b_j(c-a_j)} \period \ee
Then from the above observations about the star-triangle relation
\be \label{Mxi}
M_j \, \xi_{j+1} \eq h_j \, \xi_j \comma \ee
where 
\bd h_j = h(a_j, a_{j+1}, b_{j+1}, b_j) \ed
is a scalar factor dependent (as are $M_j$ and $\xi_j$) 
on the lower and upper spins $a_j, a_{j+1}, b_j, b_{j+1}$.
In fact
\be \label{defh}  h(a,b,c,d) \eq  \sum_g W_{6v} (v| a,b,g,f)
 W_Q(v|f,g,c,d) x^{c(g-b)-d(f-a)} \period \ee
Here $W_{6v}$, $W_Q$ are defined by (\ref{defWsv}), (\ref{WQ}). 
The rhs is necessarily independent of the extra spin $f$.

A direct calculation of (\ref{defh}) reveals that
\be 
h(a,b,c,d) \eq \sinh \half (\lambda-v) \; 
W_Q(v+2 \lambda'| a,b,c,d) \ee

 Since
$W_1$ in (\ref{defM}) is the six-vertex model weight function,
$c$ is $a_j \pm 1$ and $d$ is $a_{j+1} \pm 1$. Hence each $M_j$ is a
two-by-two matrix. Now take $P_j$ to be a two-by-two 
matrix whose first column is $\xi_j$. The second column can be chosen
arbitrarily, so long as the choice makes $P_j$ invertible. It is
convenient to choose ${\rm det}(P_j) = 1$.

Then (\ref{Mxi}) ensures that
\be\label{sim}
 P_j^{-1} M_j P_{j+1} \eq  {\tilde{M}}_j \comma \ee
where ${\tilde{M}}_j$ is an upper-triangular two-by-two matrix
with upper-left entry $h_j$. Equating determinants, we find that
the lower-right entry of  ${\tilde{M}}_j$ is 
$h'(a_j, a_{j+1}, b_{j+1}, b_j)$, where
\be 
h'(a,b,c,d) \eq \sinh \half (\lambda+v) \; 
W_Q(v-2 \lambda'| a,b,c,d) \period  \ee

Again, we have to worry about our skewed boundary conditions.
In general we do not have $P_{L+1} = P_1$, but rather
$P_{L+1} = {\cal D} P_1$, where $\cal D$ is a diagonal two-by-two
matrix with entries $x^{t(c-a_1)} \delta (c,d)$.
As in (\ref{modcommn}), we must therefore replace $T_1$ by 
$D^{-s} T_1 D^s$, which is equivalent to replacing the rhs of
(\ref{trace}) by ${\rm Trace} \, {\cal D} M_1 \cdots M_L$. Then
the similarity transformation (\ref{sim}) reduces the matrix
product to upper-triangular form, so the trace becomes
the sum of two products, the first over all upper-left elements
of the $\tilde{M}_j$, the second over all lower-right.
We obtain
\be \label{TQreln}
D^{-s} \, T^{r} \! (v) \, D^s \, Q_R^{\rs}(v)
 \eq \phi(\lambda \minus v) \, 
Q_R^{\rs} (v \plus 2 \lambda') + 
\phi(\lambda \plus v) \, Q_R^{\rs} (v \minus 2 \lambda') \comma 
\ee
where
\be \phi(v) = [\sin (v) ]^L \period \ee

\subsection*{$Q, Q$ commutation.}

Now we consider the conditions for two transfer matrices
$Q_R^{\rs}(v)$, with different values of $v$ and possibly
different values of $r, s$, to commute. This is the step at
which one first encounters the chiral Potts model.

As we have seen, the six-vertex model transfer matrices $T$ satisfy
the commutation relation (\ref{svcommn}), and the $Q$ matrices
satisfy the commutation relation (\ref{TQcommn}) with $T$. At first 
sight this would appear to imply a corresponding commutation relation
between the $Q$'s themselves. If the $T$ have unique eigenvalues,
this is so.

However, Fabricius and McCoy have emphasized that for the special
case when our parameter $x$ is a complex root of unity, then the 
eigenvalues of the $T$ matrices are degenerate, and interesting
algebraic structures emerge.\footnote{The eigenvectors of $T$ 
are then of course not unique, so it is not surprising that
(\ref{TQreln}) then defines the eigenvalues of
 $T$ uniquely, but not those of $Q$ - a fact that
Fabricius and McCoy seem to find remarkable.} It is therefore 
desirable to establish the $Q, Q$ commutation relations
directly.

We immediately strike a difficulty with our spin language. Up to now
we have only considered matrix products that involve at least one
six-vertex $T$ matrix, e.g $T Q_R$. Each element of
this product corresponds
to two rows of faces of the lattice, the lower with transfer
matrix $T$, the upper with $Q_R$. There are three rows of spins
on sites, those in the lowest and uppermost are given, the ones in 
the middle row are to be summed over. Since at least one of the 
matrices is $T$, the ice rule (all horizontally {\em and} vertically
adjacent spins differ by one) ensures that that there are a finite
number of terms in this summation.

However, if both matrices are $Q_R$ matrices, one can choose one of 
the spins in the middle row arbitrarily. At this stage we are 
allowing spins to take all integer values, so there are an infinite 
number of choices and an infinite number of terms in the summation.
The matrix product is not defined.

To overcome this, we anticipate the next section, where we go
from spin language back to ``horizontal spin difference'' or
``vertical arrow'' language. If we can arrange a $Q Q$ product
so that the terms in the sum it represents are unchanged by
incrementing all the spins in the middle row by unity, then
we can write this sum as a sum over only the differences of such 
spins. This is equivalent to arbitrarily fixing one of the 
spins in the middle row, say the left-hand spin.

{From} (\ref{WQ}) and (\ref{deftQ}), incrementing by unity all the 
spins in the lower row of the matrix $Q_R^{rs}$ leaves the 
elements of that matrix unchanged. It is not so the upper row:
that introduces a factor $x^{-r}$ coming from the 
$\exp [\lambda' d (a-b)/2] $ term in (\ref{WQ}). We can overcome this
by instead using the matrix 
\be \label{defhatQ}
   Q_L^{\rs} (v) \eq  x^{rs} D^{s} Q_R^{\rs}(v) D^r  \ee
which has an extra factor $x^{r\sigma'_1 +s \sigma_1 + rs}$ in 
(\ref{deftQ}). This is in turn equivalent to multiplying
(\ref{WQ}) by $x^{bc - ad}$, i.e. to replacing $\lambda'd (a-b)/2$
by $\lambda' b (c-d)/2$. Thus
\be \label{deftQhat}
[Q_L^{\rs}(v) ]_{\sigma, \sigma'} \eq 
\hat{W}_Q(v|\sigma_1,\sigma_2, \sigma'_2,\sigma'_1) \cdots  
\hat{W}_Q(v|\sigma_L,\sigma_{L+1}, \sigma'_{L+1} ,\sigma'_L) \comma 
\ee
where
\be \label{WQhat}
\hat{W}_Q(v|a,b,c,d) \eq \exp[{\lambda' b (c-d)/2}  - 
(\lambda+v)(b-a)(c-d)/4] \period \ee
This matrix $Q_L$ corresponds to the $Q_L$ of (9.8.27) of 
Ref. \cite{book82}, but with the normalization changed and
both $\lambda$ and $v$ negated.
 
This matrix $Q_L^{\rs}$ is unchanged by incrementing all the 
upper spins by unity, and we can define the product
\be \label{defI}
 I^{rts}(v,v') = Q_L^{\rt}(v) \, Q_R^{\ts}(v') \ee as above.

Let $\{ b_1, \ldots , b_{L+1} \}$ be the spins that are summed over
in this matrix product, i.e. the spins on the middle row, above 
$ Q_R^{rt}(v) $ and below $Q_L^{\ts}(v') $. They enter
only via their differences $\beta_j = b_{j+1} - b_j$. The sum is over  
all values ($\pm 1$) of the  $\beta_j$, subject to the skew condition
\bd \beta _1 + \cdots + \beta_L \eq t \period \ed
We can remove this restriction by forming the generating function 
matrix
\be \label{deftau}  F^{rs}(z|v,v')  \eq \sum_t z^t I^{rts}(v,v') \ee
the sum being over all values of $t$ from $-L$ to $L$, with
$L-t$ even. Then $F^{rs}(z|v,v') $ is given by the unrestricted
sum over the $\beta_j$.

Being a transfer matrix product, the elements of
$F^{rs}(z|v,v') $ are given in the first instance 
by an expression of type (\ref{trace}). However, now the elements
$[M_j]_{c,d}$ of $M_j$ depend on the two intermediate spins $c,d$ 
only via their difference $m = d-c = \pm 1$. The trace in (\ref{trace})
simplifies to a sum over all the $L$ spin differences. These are now 
independent, so we can perform the $m$ summation individually
for each $M_j$. The rhs of (\ref{trace}) then becomes a product,
as in (\ref{defT}). Hence
the elements of $ F^{rs}(z|v,v') $ are given by the rhs of 
(\ref{defT}), with $W$ replaced by $W_{F}$, and 
\be \label{defWcomb}
W_{F}(v,v'|a,b,c,d) = \sum_m z^m \hat{W}_Q(v|a,b,f+m,f)
 W_Q(v'|f,f+m,c,d ) 
\comma \ee
the sum being over $m = -1$ and $+1$.  The rhs is independent of $f$
and from (\ref{WQ}), (\ref{WQhat})
\be \label{defWtau}
W_{F}(v,v'|a,b,c,d) = \sum_m z^m x^{m(b-d)} \, 
{\rm e}^{m[(\lambda+v)m (a-b) +
(\lambda+v') (d-c)]/4} \period \ee
Here
$v, v'$ are independent variables.

We look for an auxiliary function $P(v,v'|a,b)$ such that
\be \label{QQcr} 
W_{F}(v',v|a,b,c,d) \eq P(v,v'|d-a) W_{F}(v,v'|a,b,c,d)/P(v,v'|c-b) 
\comma \ee
since then (apart possibly from boundary conditions) the rhs of 
(\ref{defT}) is unaltered by interchanging $v$ with $v'$, as the $P$
factors cancel out of the product on the rhs. This
in turn implies a commutation relation between
 $Q_L^{\rt}(v)$ and $ Q_R^{\ts}(v')$.

There are four cases to consider in (\ref{QQcr}),
$b-a = \pm 1$ and $c-d = \pm 1$. If $b-a = c-d$, then (\ref{QQcr}) is 
automatically satisfied. The other two cases yield just one
condition on the function $P$, which can be written
\be \label{defP}
\frac{P(v,v'|j+1)}{P(v,v'|j-1)} \eq \frac{z \, {\rm e}^{2v}
 + x^{2j} {\rm e}^{2v'}}{z \, {\rm e}^{2v'}
 + x^{2j} {\rm e}^{2v}} \period \ee
Hence $P(j/2)$ has the same structure as the chiral Potts weight 
function $W_{pq}(j)$ in eqn (2) of \cite{BPAY88}, with 
$\omega = x^4$. However, we emphasize that at this stage we are 
{\em not}  requiring $x^4$ to be a root of unity, so the rhs of 
(\ref{defP}) is not necessarily periodic in $j$.

Substituting the form (\ref{QQcr}) for $W = W_{F}$ into 
(\ref{defT}), the $P$-functions cancel except
for a boundary factor
$P(v,v'|\sigma'_1 \minus  \sigma_1)/P(v,v'|\sigma'_{L+1} 
\minus \sigma_{L+1})$.
Remembering that $\sigma_{L+1} = \sigma_1 + r$ and
$\sigma'_{L+1} = \sigma'_1 + s$, this has the form 
$P(v,v'|d \minus a)/P(v,v'|d \minus a \plus s \minus r)$. 
This is unity  if $s = r$, so we have established the symmetry
property  ${F}^{rr}(z|v',v) = {F}^{rr}(z|v,v') $. This is true for
{\em all} $z$, so the coefficients $I^{rtr}(v,v') $ in
the Laurent expansion of ${F}^{rr}(z|v',v)$ are also symmetric
in $v,v'$, for all $t$. Hence from (\ref{defI})
\bd Q_L^{\rt}(v) \, Q_R^{\tr}(v')  \eq 
Q_L^{\rt}(v') \, Q_R^{\tr}(v)  \period \ed
{From} (\ref{defhatQ}), replacing $t$ by $s$, this becomes
\be \label{QQcommn}
 Q_R^{\rs}(v) D^r Q_R^{\sr}(v') \eq 
Q_R^{\rs}(v') D^r Q_R^{\sr}(v) \period \ee

\section{SUMMARY OF THE RELATIONS}

Neither $Q_R^{\rs}(v)$ nor $Q_L^{\rs}(v)$ has the desirable 
property that it is unchanged by incrementing all the spins in 
{\em either } the lower or the upper row in Figure \ref{row} by 
unity. {From} the discussion preceding (\ref{defhatQ}), the matrix
that does have this property is
\be \label{defQ}
 \oQ^{rs}(v) \eq Q_R^{\rs}(v) D^r \eq x^{-rs} \, D^{-s} Q_L^{\rs}(v) \period \ee

It is convenient to introduce a generalization of the six-vertex 
model  transfer matrix $T^r(v)$:
\be \label{defTrs}
T^r_s(v) \eq D^{-s} T^r(v) D^s  \period \ee
This is the transfer matrix of the six-vertex model with skew 
parameter $r$ and a field of weight $x^{\sigma'_1 - \sigma_1}$ 
acting on the two vertically adjacent spins (or equivalently 
the horizontal arrow between them) at the left-hand end of the row.
Then the commutation and functional relations (\ref{svcommn}),
(\ref{TQcommn}), (\ref{TQreln}), (\ref{QQcommn})    for the 
$T$ and $Q$  matrices become
\be \label{svcommn2}
T^r_s(v) \, T^r_s(v') \eq T^r_s(v') \, T^r_s(v) \comma \ee

\be \label{TQcommn2}
 T^{r}_s \! (v) \, \oQ^{rs}(v') \eq 
\oQ^{rs}(v') \, T^{s}_r(v) \comma \ee

\be \label{TQreln2}
 T^{r}_s \! (v) \,  \oQ^{rs}(v)
 \eq \phi(\lambda \minus v) \, 
\oQ^{rs} (v \plus 2 \lambda') + 
\phi(\lambda \plus v) \, \oQ^{rs} (v \minus 2 \lambda') \comma \ee

\be \label{QQcommn2}
   \oQ^{rs}(v) \oQ^{sr}(v') \eq 
\oQ^{rs}(v') \oQ^{sr}(v)  \ee

\noindent for all values of the skew parameters $r, s$. Note that 
nowhere in this paper do we use any implicit summation convention
over skew parameters.

\subsection*{Square matrix form of the relations}

We have uncovered quite an interesting structure
in the $T,Q$ relations for the six-vertex model, and we could 
continue to work with the above relations (\ref{svcommn2}) - 
(\ref{QQcommn2}). However, they are still not quite in the
form of the  $T,Q$ (or $V, Q$) relations given in 
(9.2.5), (9.8.40), (9.8.42) of Ref. \cite{book82}. These involve
the zero-field six-vertex matrix $T^r(v)$ and a square invertible 
matrix $Q(v)$, for given $r$.

To derive these square matrix relations,  define, using 
(\ref{QQcommn2}),
\be \label{deffinalQ}
Q^r_s(v) = \oQ^{rs}(v)  \, \oQ^{sr} (v_0)  = 
 \oQ^{rs}(v_0)  \, \oQ^{sr} (v) \ee
$v_0$ being some fixed value of $v$. 

Then from (\ref{svcommn2}) - (\ref{QQcommn2}) we readily obtain
the desired final square matrix relations
\bd T^r_s(v) \, T^r_s(v') \eq  T^r_s(v')  \,  T^r_s(v)  \comma \ed
\be \label{final}
 T^r_s(v)  \, Q^r_s  (v') \eq  Q^r_s  (v')  \, T^r_s(v)  
\comma \ee
\bd 
 T^{r}_s \! (v) \,  Q^{r}_s  (v)
 \eq \phi(\lambda \minus v) \, 
Q^{r}_s   (v \plus 2 \lambda') + 
\phi(\lambda \plus v) \, Q^{r}_s   (v \minus 2 \lambda') \comma \ed
\bd 
   Q^{r}_s  (v)  \, Q^{r}_s (v') \eq 
Q^{r}_s  (v')  \, Q^{r}_s (v)  \comma \ed
true for all the allowed values $ -L , 2-L , 4-L,  \ldots , L$ of $r$
and $s$.

{From} our remarks in the next sub-section,  we expect $Q^r_s(v)$ 
to be invertible provided $C_s \geq C_r$. For an even number $L$ 
of columns, this will always be so if we choose
$s=0$. Then $T^r_0(v) = T(v)$ is the usual zero-field six-vertex 
model transfer matrix and $Q^r_0(v) = Q(v)$ is the associated 
$Q$-matrix used in \cite{Baxter71} - \cite{book82} . We have 
derived the relations
(\ref{TTcommn}) - (\ref{QQcommna}), except that they can be and 
usually are thought of in the ``arrow language'' of the next 
sub-section.

\subsection*{Transformation to arrow language}

We have used the language of spins on sites to derive these relations, 
but the $T, Q$ matrices can be thought of conveniently
in the alternative spin-difference or ``arrow'' language, 
transforming from the 
$\sigma_j$ to the $\alpha_j$ by (\ref{defalpha}). The equations
(\ref{defD}), (\ref{WQ}), (\ref{deftQ}), (\ref{defQ}) define
the elements $(\sigma, \sigma')$ of $\oQ^{rs}$. They depend only on
$(\alpha, \alpha')$, so we take the elements to be labelled by
$\alpha$, $\alpha'$. Since the $\alpha_j$ satisfy (\ref{sumalphas}),
and the  $\alpha'_j$ satisfy (\ref{sumalphas}) with $r$ replaced by 
$s$, the matrix $\tilde{Q}^{rs}$ is of dimension $C_r$ by $C_s$, where
\bd
C_r \eq  \left( \begin{array}{c} L \\ (L+r)/2 \end{array} 
\right)    \ed
and $Q^r_s(v)$ is a square matrix, of dimension $C_r$.

{From} this point of view the six-vertex model transfer matrix is 
slightly more subtle, since we are not free to increment all the 
spins in the upper row of Figure \ref{row} independently of those
in the lower row. Only the face configurations of Figure 
\ref{sixvertex} are allowed. If the spins in the lower row
 are known , then there two choices for those in the upper, 
depending on whether $\sigma'_1 = \sigma_1 +1$ or 
$\sigma'_1 = \sigma_1 -1$. However, the above
matrix products involving $T$ always involve a sum over these two 
choices. The end result is that the above relations 
(\ref{svcommn2}) - (\ref{QQcommn2}) are also true in the 
arrow language provided we sum over the two choices of $\sigma'_1$.
{From} (\ref{defD}), (\ref{defT6v}), (\ref{defTrs}) we obtain 
\be[ T^{r}_s \! (v) ]_{\alpha, \alpha'} \eq \sum_{\beta} 
x^{s \beta} W_{6v}(v|\sigma_1,\sigma_2,\sigma'_2,\sigma'_1 ) 
 \ldots  W_{6v} (v|\sigma_L,\sigma_{L+1},\sigma'_{L+1},\sigma'_L )  
\comma \ee
the sum being over $\beta = \pm 1$, where 
$\beta = \sigma'_1 - \sigma_1$. The rhs is uniquely defined by
$\alpha_1, \ldots \alpha'_{L}$. This is the usual six-vertex model
transfer matrix in arrow language, with an extra weight $x^s$ 
($x^{-s}$) if the left-hand horizontal arrow points to the 
left (right). It is of dimension $C_r$ by $C_r$.

\subsection*{Transformation to all $T^r_s(v)$ diagonal}

We assume that the matrices $T^r_s(v), \, Q^r_s(v)$ are 
diagonalizable. Then from the set of relations (\ref{final}), 
there exists a non-singular  matrix ${\cal P}^r_s$, independent 
of $v$, such that the transformation 
$T^r_s(v) \rightarrow {\cal P}^r_s T^r_s(v) \{ {\cal P}^r_s \}^{-1}$ 
reduces both $T^r_s(v) $ and $Q^r_s(v)$  to diagonal form. If we 
also  transform $\oQ^{rs}(v)$
by $\oQ^{rs}(v) \rightarrow {\cal P}^r_s \oQ^{rs}(v) 
\{ {\cal P}^s_r \}^{-1}$,
then the relations (\ref{svcommn2}) - (\ref{QQcommn2}) and 
(\ref{final}) are unchanged.

We do not have a proof for general $L$, but for $L = 2$ to $L= 6$
we have verified with Mathematica that the matrix product
$\oQ^{rs}(v) \oQ^{sr}(v')$ is invertible if $C_s \geq C_r$, for all 
such values of $r$ and $s$ and for arbitrarily chosen
 values of $v, v', x$. This implies that the rectangular matrix
$\oQ^{rs}(v)$ is then of ``full'' rank, having the maximum rank 
possible for a matrix of its size:
\be \label{rankQ}
{\rm rank \; \; of \; \; } \oQ^{rs} \eq \min (C_r, C_s) \period \ee

{From} the transformed form of (\ref{TQcommn2}) this implies that 
if $C_s \geq C_r$, then the eigenvalues of $T^r_s(v)$ are all 
contained in the eigenvalues of $T^s_r(v)$, which is an interesting 
result of which the author was previously not fully aware. If the 
eigenvalues of $T^s_r(v)$ are non-degenerate, 
(\ref{TQcommn2})  further implies that one can order the
eigenvalues so that $\oQ^{rs}(v)$ transforms to a 
diagonal (but not necessarily square) matrix. Even if they are 
degenerate, then the fact that the transform of $Q^r_s(v)$ is also 
diagonal may imply this property of $\oQ^{rs}(v)$.

\subsection*{Structure of the eigenvalues}

 If $C_s >= C_r$, then  it follows  from (\ref{WQ}) and 
(\ref{deftQ}) that for $r >= 0$ all
elements of $Q_R^{\rs}(v)$ are of the form
\bd {\rm e}^{v(r-s-L)/4} \; \times \{ \; {\rm  polynomial \; \; in 
\;\;} {\rm e}^v \; \; {\rm of \; \; degree \; \; } (L-r)/2 \; \} 
\semicolon \ed
%%3456789012345678901234567890123456789012345678901234567890123456789012
while for $r<= 0$  they are of the form
\bd {\rm e}^{v(s-r-L)/4} \; \times \{ \; {\rm  polynomial \; \; 
in \;\;} {\rm e}^v \; \; {\rm of \; \; degree \; \; } (L+r)/2 \; \} 
\period \ed
If $C_s <= C_r$, then the above statements are true if one interchanges 
$r$ with $s$ therein.
(All four statements are actually true for all $r$ and $s$, 
provided one allows 
some of the initial and/or final coefficients of the polynomial to be 
zero,  which means that some of the $v_j$ below will be infinite.)

{From} (\ref{defQ}) and (\ref{deffinalQ}), the same must be true  for 
the elements of $Q^r_s(v)$. Also, since the eigenvector
matrices ${\cal P}^r_s$ are independent of $v$, it is true of the 
diagonalized form of  $Q^r_s(v)$, i.e. of its eigenvalues. If
$C_s >= C_r$ we can  therefore write any eigenvalue as 
\bd
C \, {\rm e}^{- v s/4} \prod_{j=1}^{(L-r)/2} \sinh \half (v-v_j) \ed
if $r >= 0$, or
\bd
C \, {\rm e}^{ v s/4} \prod_{j=1}^{(L+r)/2} \sinh \half (v-v_j) \ed
 if $r <= 0$. Here $C, v_1, v_2, \ldots $ are constants, independent 
of $v$.

Similarly, the eigenvalues of $T^r_s(v)$ necessarily have the structure
\bd C' \, \prod_{j=1}^L  \sinh \half (v-w_j) \comma \ed
 $C', w_1, w_2, \ldots $ being other  constants.

We can substitute these forms into (\ref{final}) and in principle solve 
for the various constants. Setting $v = v_j$ gives the usual Bethe 
ansatz  equations. Alternatively, one can expand the functions as 
polynomials in ${\rm e}^v$ and equate coefficients.

%%3456789012345678901234567890123456789012345678901234567890123456789012

\section{A MORE GENERAL FORM FOR $Q$}

We remark that (\ref{defWtau}) is
the Boltzmann weight of the ``$\tau_2$'' model rotated through 
$90^{\circ}$ - eqns.  (3.44), (3.48) of Ref. \cite{BBP90}.
If instead of (\ref{TQstartri}) we take
\bd
W_1 = W_{6v}(v) \sep W_2 = W_{F}(v',v'') \sep 
 W_3 = W_{F}(v'-v-\lambda,v''-v-\lambda )  \comma \ed
then the star triangle relation 
(\ref{startri}) remains  satisfied, for all $v, v', v''$.

  This is the quite general solution of (\ref{startri}) found 
in 1990 by Bazhanov and Stroganoff.\cite{BazStrog90} There are 
further trivial factors that can be introduced into the Boltzmann 
weight $W_{F}(v,v'|a,b,c,d)$ without affecting
(\ref{startri}), notably $\mu_1^{b-a} \mu_2^{c-d}$, 
where $\mu_1, \mu_2$ are
arbitrary constants (the same for both $W_2$ and $W_3$). This 
factor merely multiplies 
the transfer matrix ${F}^{rr}(z|v,v')$ by $\mu_1^r \mu_2^r$.

{From} (\ref{defhatQ}), (\ref{defI}), (\ref{deftau}), (\ref{defQ}) and
the discussion before (\ref{QQcommn}),
\bd 
{F}^r(v,v') = {F}^r(v',v)  = {F}^{rr} (z|v,v') = 
\sum_t z^t x^{rt} D^t \oQ^{rt} (v) 
\oQ^{tr} (v') D^{-t} \period \ed
It does indeed follow from (\ref{defTrs}), (\ref{TQreln2}) that
\bd
T^r(v) \,  {F}^r(v',v'') \eq {F}^r(v',v'')\,  T^r(v) \comma \ed
for all $v, v', v''$. This is in agreement with the fact that 
$W_{F}(v',v'')$ satisfies (\ref{startri}). We stress that this 
is a ``spin language'' result: ${F}^r(v',v'')$ is {\em not}
unchanged by incrementing all the spins in either the lower 
or upper row of Figure \ref{row} by unity.

\section{CONCLUSION}

We have obtained the ``$T,Q$'' functional matrix relations
for the zero-field six-vertex model (and for other special 
values of the horizontal field), presenting the working in a  
way that we hope is more explicit and transparent than 
those used previously. As a check on our reasoning, we have
verified the relations (\ref{svcommn2}) - (\ref{QQcommn2})
explicitly on a computer, using Mathematica, for $L = 2$ to 
$L = 6$.

In some ways this method appears to be less general than
Lieb and Sutherland's original Bethe ansatz solutions: it is
not obvious how to extend it to the more general situation
when there is an arbitrary  horizontal electric field
(which corresponds to a vertical spin field), or even to solve the 
zero-field problem with an odd number of columns. Still, it should
be remembered that the commuting transfer matrix method
was  originally developed as a means of solving the eight-vertex 
model, for which there was then no Bethe ansatz.

We emphasize that {\em all} the above working is for 
the general case, when there are no restrictions on the
real or complex ``crossing parameter'' $\lambda$.

\subsection*{The ``root of unity'' cases.}

A very interesting special case occurs when $x^4$ is an $N$th root 
of unity, i.e.
\be \lambda' = {\rm i} m \, \pi/N \sep x^{4N} = 1 \comma \ee
$m, N$ being  integers and $k$ being odd. These cases have been studied 
by Fabricius and McCoy, and they have shown that the eigenvalues of
$T^r(v)$ are then degenerate.\cite{DegFabMcCoy01} - \cite{FabMcCoy01c}

One can then 
regard the lattice spins $\sigma_j$, $\sigma'_j$ as restricted
to $N$ values. For a given site these will either be odd or even,
e.g. $1,3, \ldots , 2N-1$. Horizontally adjacent spins differ by 
$\pm 1$ modulo $2 N$, so we continue to take the $\alpha_j$ in 
(\ref{defalpha}) to be strictly $\pm 1$, but interpret the 
difference on the rhs to modulo $2 N$. Similarly,
the skew parameter $r$, given by (\ref{defskewparm}), is to be
interpreted modulo $2 N$. One can then repeat all the above working.
(There may have to be some minor modifications, such as multiplying
$Q_R^{\rs}(v)$ by ${\rm e}^{{\rm i} \pi r/N}$ so as to 
ensure that incrementing the spin $d$ in (\ref{WQ}) by $2 N$
leaves the element of $Q_R^{\rs}(v)$ unchanged. The 
horizontal spin differences $a-b$, $c-d$ therein are to be kept
as strictly $\pm 1$.) 

There will be one significant difference: in the above we have
specialized to the case when no horizontal spin fields 
(corresponding to vertical arrow or ``electric'' fields) are 
applied to the lower or upper rows of the $T$ or $Q$ matrices, 
for instance including a factor $\mu^{b-a}$ in (\ref{WQ}). We 
could have introduced such a factor as
it does not affect the star-triangle relation (\ref{startri}), 
(\ref{TQstartri}). It merely post-multiplies  
$Q_R^{\rs}(v)$ by the diagonal matrix $S^z$ defined in 
section 3.
Since  $S^z$ commutes with $T^r_s(v)$ and $Q^r_s(v)$, the 
equations (\ref{final}) would have been unchanged: no increase 
in generality  would have been achieved.   

For the root of unity cases this ceases to be true:
$\mu^r$ is {\em not}  in general the same as $\mu^{r+2N}$.
$S^z$ now commutes with $T^r_s(v)$, but {\em not} with
$Q^r_s(v)$.  
For generality it is therefore important to include such factors
in (\ref{WQ}). This leads to an extra constant term $\cal C$
on the rhs of (\ref{defP}). {From} (\ref{QQcr} ),
remembering that $|a-b| = |c-d|= 1$, the function $P(v,v'|j)$ is
defined only for either $j$  always odd, or $j$ always even.
Taking $j$ to be odd, from (\ref{defP}),
\be \label{chirP} 
 P(v,v'|2n-1) \eq \prod_{j=1}^n {\cal C} \, 
\frac{z \, {\rm e}^{2v}
 + \omega^{j} {\rm e}^{2v'}}{z \, {\rm e}^{2v'}
 + \omega^{j} {\rm e}^{2v}} \comma \ee
where
\bd \omega = x^4  \sep \omega ^N = 1\period \ed

The function $P(v,v'|j)$ must now be periodic in $j$ of period $2N$,
which will be so iff
\bd {\cal C}^N \{ (-z)^N {\rm e}^{2Nv} -{\rm e}^{2 N v'} \} \eq
 (-z)^N {\rm e}^{2Nv'} - {\rm e}^{2 N v} \period \ed

With a correct identification of  the 
various parameters, we see that \\ $P(v,v'|2n-1)$ is precisely 
the $N$-state chiral Potts model
weight function $W_{pq}(n)$.\cite{BPAY88} Further, the factor
$\cal C$ arising from the applied fields  is important: without it
one would have $z = {\rm e}^{{\rm i} \pi/N}$ and the model would 
reduce to the critical Fateev-Zamolodchikov model.\cite{FatZam82}

\subsection*{Kashiwara-Miwa model as a descendant 
of the eight-vertex model}

Just as the chiral Potts model can be regarded as a ``descendant''
of the six-vertex model in a vertical arrow field, so Hasegawa and 
Yamada have shown that the Kashiwara-Miwa model can be regarded
as a ``descendant'' of the zero-field eight-vertex 
model.\cite {HasYam90, KashMiwa86} Note that neither vertex model 
includes the other: they  intersect in the zero-field 
six-vertex model. Correspondingly,  the chiral Potts and Kashiwara 
model intersect in the Fateev-Zamolodchikov model.

In the same way that our present approach leads very directly 
from the six-vertex model to the chiral Potts model, so does it lead 
from the eight-vertex model to the Kashiwara-Miwa model.
For the eight-vertex model, the function $W_F(u,v|a,b,c,d) $
is the multiplicand of   eqn. (10.5.27) of \cite{book82}, the 
$s_j, s'_j, \sigma_j, \sigma'_j$ therein being
\be
s_j = s + \lambda d \sep s'_j = s' + \lambda a \sep 
\sigma_j = c-d \sep \sigma'_j = b-a \comma \ee
and $s, s'$ are arbitrary parameters.\footnote{Eqn. (10.5.8) 
of \cite{book82} contains an error: 
$\sigma_{j+1}$ therein should be $\sigma_{j-1}$.}

Here we established the $Q, Q$ commutation by showing that
$F^{rr}(z|v,v') = F^{rr}(z|v',v)$. Similarly, in \cite{book82} the 
author established the relation (10.5.29) by showing that the product
(10.5.27) is a symmetric function of $v$ and $v'$. That argument was
by inductive reasoning on the entire product. If instead we look
for a solution for the function $P$ in (\ref{QQcr}), we
find that the $P$-factors depend on $a, d$ (or $b, c$) not just via their 
difference, but now $P(v,v'|a,d)$ is a product of a function of $d-a$ 
difference and a function of $a+d$:
\bd P(v,v'|a,d) =  P_1(v,v'|d-a) P_2(v,v'|a+d) \period \ed
Focussing on the root of unity case, when 
$\lambda' = \lambda - 2 \rii K = 2 {\rm i} m \, K/N$, $m$ and $N$ being 
integers, and requiring that 
$P(v,v'|a+2N,d) = P(v,v'|a,d+2N) = P(v,v'|a,d)$,
we find that $P(v,v'|a,d)$ is the Boltzmann weight function of the
Kashiwara-Miwa model. Since the eight-vertex model transfer matrix
does not commute with $S^z$, it does not break up into diagonal blocks
with  different $r$ and there is no analogue of the arbitrary factor 
$\cal C$ in  (\ref{chirP}).

%%3456789012345678901234567890123456789012345678901234567890123456789012

\newpage

%% Figure 1

\setlength{\unitlength}{1pt}
\begin{figure}[hbt]
\begin{picture}(420,140) (-25,0)
\put(32,114) {Six-vertex model}
\put(45,100) {in a field}
\put(28,90) {\line(1,0) {95}}
\put(28,130) {\line(1,0) {95}}
\put(28,90) {\line(0,1) {40}}
\put(123,90) {\line(0,1) {40}}

\put(157,114) {Zero-field}
\put(157,100) {six-vertex}
%%\put(190,90) {\line(1,0) {60}}
%%\put(190,130) {\line(1,0) {60}}
%%\put(190,90) {\line(0,1) {40}}
%%\put(250,90) {\line(0,1) {40}}
\multiput(140,90)(3,0){28}{.}
\multiput(140,130)(3,0){28}{.}
\multiput(140,90)(0,3){14}{.}
\multiput(223,90)(0,3){14}{.}
\put(224,110) {\line(1,0) {19}}
\put(123,110) {\line(1,0) {17}}
\put(224,109) {\line(1,0) {19}}
\put(123,109) {\line(1,0) {17}}

\put(253,114) {Zero-field}
\put(248,100) {eight-vertex}
\put(243,90) {\line(1,0) {70}}
\put(243,130) {\line(1,0) {70}}
\put(243,90) {\line(0,1) {40}}
\put(313,90) {\line(0,1) {40}}

\put(75,40) {\line(0,1) {50}}
\put(74,40) {\line(0,1) {50}}
%%\put(75,80) {\vector(0,-1) {20}}
\put(75,58) {\line(1,2) {5}}
\put(75,58) {\line(-1,2) {5}}
\put(42,17) {Chiral Potts}
\put(28,0) {\line(1,0) {95}}
\put(28,40) {\line(1,0) {95}}
\put(28,0) {\line(0,1) {40}}
\put(123,0) {\line(0,1) {40}}

\put(180,40) {\line(0,1) {50}}
\put(181,40) {\line(0,1) {50}}
\put(181,58) {\line(1,2) {5}}
\put(181,58) {\line(-1,2) {5}}
\put(158,24) {Fateev-}
\put(145,10) {Zamolodchikov}
%%\put(190,0) {\line(1,0) {60}}
%%\put(190,40) {\line(1,0) {60}}
%%\put(190,0) {\line(0,1) {40}}
%%\put(250,0) {\line(0,1) {40}}
\multiput(140,0)(3,0){28}{.}
\multiput(140,40)(3,0){28}{.}
\multiput(140,0)(0,3){14}{.}
\multiput(223,0)(0,3){14}{.}
\put(224,20) {\line(1,0) {19}}
\put(123,20) {\line(1,0) {17}}
\put(224,19) {\line(1,0) {19}}
\put(123,19) {\line(1,0) {17}}

\put(277,40) {\line(0,1) {50}}
\put(278,40) {\line(0,1) {50}}
\put(278,58) {\line(1,2) {5}}
\put(278,58) {\line(-1,2) {5}}
\put(248,24) {Kashiwara-}
\put(258,10) {Miwa}
\put(243,0) {\line(1,0) {70}}
\put(243,40) {\line(1,0) {70}}
\put(243,0) {\line(0,1) {40}}
\put(313,0) {\line(0,1) {40}}

\end{picture}
\vspace{0.5cm}
\caption{ The three vertex models and their respective ``descendants''. The central models 
are special cases of the ones to their left and to their right.}
\label{models}
\end{figure}
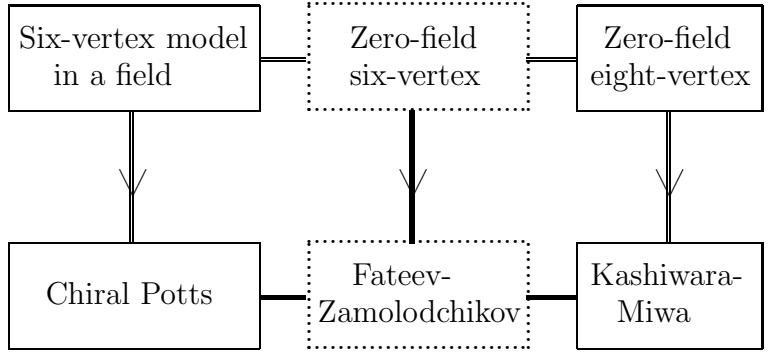

\vspace{5cm}

%% Figure 2

\setlength{\unitlength}{1pt}
\begin{figure}[hbt]
\begin{picture}(420,100) (-70,0)
\put(80,0) {\line(1,0) {60}}
\put(80,60) {\line(1,0) {60}}
%%\multiput(50,30)(5,0){24}{.}
%%\multiput(110,-10)(0,5){19}{.}
%%\put(110,-20) {$p$}
%%\put(171,27) {$q$}
\put(80,0) {\line(0,1) {60}}
\put(140,0) {\line(0,1) {60}}
\put (72,-11) {$a$}
\put (140,-11) {$b$}
\put (72,66) {$d$}
\put (140,66) {$c$}
\end{picture}
\vspace{0.5cm}
\caption{ Four spins round a face of the square lattice. }
\label{fourspins}
\end{figure}
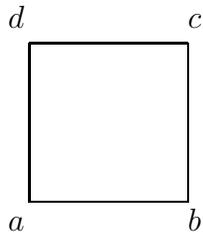

\newpage

%% Figure 3

\setlength{\unitlength}{1pt}
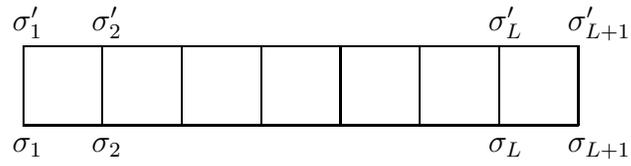
\begin{figure}[hbt]
\begin{picture}(420,80) (-40,0)
\put (60,0) {\line(1,0) {210}}
\put (60,30) {\line(1,0) {210}}
\multiput(60,0)(30,0){8}{\line(0,1) {30}}
\put (56,-10) {${\sigma_1}$}
\put (56,36) {${\sigma'_1}$}
\put (86,-10) {${\sigma_2}$}
\put (86,36) {${\sigma'_2}$}
\put (236,-10) {${\sigma_{L}}$}
\put (236,36) {${\sigma'_{L}}$}
\put (266,-10) {${\sigma_{L+1}}$}
\put (266,36) {${\sigma'_{L+1}}$}
 \end{picture}
\vspace{1.5cm}
 \caption{ A typical row of $L$ faces of the square lattice, with
spins $\sigma_1, \ldots , \sigma_{L+1}$ on the lower row of sites,
$\sigma'_1, \ldots , \sigma'_{L+1}$ on the upper.}
 \label{row}
\end{figure}  

\vspace{5cm}

%% Figure 4

\setlength{\unitlength}{1pt}
\begin{figure}[hbt]
\begin{picture}(420,120) (-40,0)
\put(40,0) {\line (1,0) {40}}
\put(20,34) {\line (1,0) {40}}
\put(80,0) {\line (3,5) {20}}
\put(20,34) {\line (3,-5) {20}}
\put(60,34) {\line (3,-5) {20}}
\put(40,68) {\line (1,0) {40}}
\put(20,34) {\line (3,5) {20}}
\put(60,34) {\line (3,5) {20}}
\put(80,68) {\line (3,-5) {20}}
\put(180,0) {\line (1,0) {40}}
\put(200,34) {\line (1,0) {40}}
\put(220,0) {\line (3,5) {20}}
\put(180,0) {\line (3,5) {20}}
\put(160,34) {\line (3,-5) {20}}
\put(180,68) {\line (1,0) {40}}
\put(160,34) {\line (3,5) {20}}
\put(220,68) {\line (3,-5) {20}}
\put(180,68) {\line (3,-5) {20}}
\put(10,31) {$a$}
\put(32,-10) {$b$}
\put(80,-10) {$c$}
\put(103,31) {$d$}
\put(80,73) {$e$}
\put(32,73) {$f$}
\put(53,26) {$g$}
\put(43,10) {$W_1$}
\put(43,47) {$W_2$}
\put(74,31) {$W_3$}
\put(150,31) {$a$}
\put(174,-10) {$b$}
\put(220,-10) {$c$}
\put(244,31) {$d$}
\put(220,73) {$e$}
\put(174,73) {$f$}
\put(201,26) {$g$}
\put(204,10) {$W_2$}
\put(204,47) {$W_1$}
\put(173,31) {$W_3$}
\put(125,31) {$=$}
 \end{picture}
\vspace{1.5cm}
 \caption{ The generalized star-triangle relation.}
 \label{startrifig}
\end{figure}
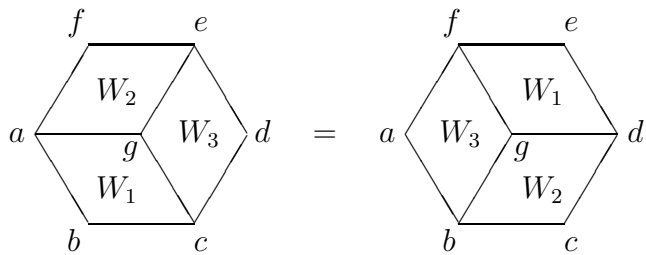

\newpage

%% Figure 5

\setlength{\unitlength}{1pt}
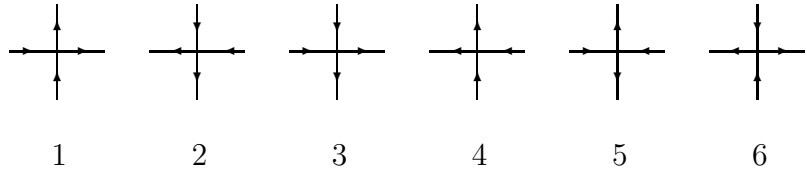
\begin{figure}[hbt]
\begin{picture}(420,80) (-40,0) 
\multiput(17,15)(53,0){6}{\line(1,0) {36}}
\multiput(35,-3)(53,0){6}{\line(0,1) {36}}
\put(18,15) {\vector(1,0) {8}}
\put(35,15) {\vector(1,0) {12}}
\put(90,15) {\vector(-1,0) {12}}
\put(106,15) {\vector(-1,0) {8}}
\put(124,15) {\vector(1,0) {8}}
\put(141,15) {\vector(1,0) {12}}
\put(196,15) {\vector(-1,0) {12}}
\put(211,15) {\vector(-1,0) {8}}
\put(230,15) {\vector(1,0) {8}}
\put(263,15) {\vector(-1,0) {8}}
\put(301,15) {\vector(-1,0) {12}}
\put(300,15) {\vector(1,0) {12}}
\put(35,0) {\vector(0,1) {8}}
\put(35,15) {\vector(0,1) {12}}
\put(88,15) {\vector(0,-1) {12}}
\put(88,30) {\vector(0,-1) {8}}
\put(141,15) {\vector(0,-1) {12}}
\put(141,30) {\vector(0,-1) {8}}
\put(194,0) {\vector(0,1) {8}}
\put(194,15) {\vector(0,1) {12}}
\put(247,15) {\vector(0,-1) {12}}
\put(247,15) {\vector(0,1) {12}}
\put(300,0) {\vector(0,1) {8}}
\put(300,30) {\vector(0,-1) {8}}
\put(33,-28) {1}
\put(86,-28) {2}
\put(139,-28) {3}
\put(192,-28) {4}
\put(245,-28) {5}
\put(298,-28) {6}
 \end{picture}
\vspace{1.5cm}
 \caption{The six arrow configurations of the six-vertex model.}
 \label{sixv}
\end{figure}

\vspace{5cm}

%% Figure 6

\setlength{\unitlength}{1pt}
\begin{figure}[hbt]
\begin{picture}(420,80) (-40,0)
\multiput(20,0)(53,0){6}{\line(1,0) {30}}
\multiput(20,30)(53,0){6}{\line(1,0) {30}}
\multiput(20,0)(53,0){6}{\line(0,1) {30}}
\multiput(50,0)(53,0){6}{\line(0,1) {30}}
\put (16,-10) {$\scriptstyle{a}$}
\put (40,-10) {$\scriptstyle{a +1} $}
\put (16,36) {$\scriptstyle{a-1}$}
\put (48,36) {$\scriptstyle{a } $}
\put(33,-28) {1}
\multiput(53,0)(53,0){1}{\put (19,-10) {$\scriptstyle{a }$}
\put (43,-10) {$\scriptstyle{a-1 } $}
\put (16,36) {$\scriptstyle{a +1}$}
\put (48,36) {$\scriptstyle{a } $}
\put(33,-28) {2}}
\multiput(106,0)(53,0){1}{\put (19,-10) {$\scriptstyle{a }$}
\put (40,-10) {$\scriptstyle{a-1 } $}
\put (14,36) {$\scriptstyle{a-1 }$}
\put (40,36) {$\scriptstyle{a-2 } $}
\put(38,-28) {3}}
\multiput(159,0)(53,0){1}{\put (19,-10) {$\scriptstyle{a }$}
\put (41,-10) {$\scriptstyle{a +1 } $}
\put (16,36) {$\scriptstyle{a +1 }$}
\put (41,36) {$\scriptstyle{a +2 } $}
\put(33,-28) {4}}
\multiput(212,0)(53,0){1}{\put (19,-10) {$\scriptstyle{a }$}
\put (40,-10) {$\scriptstyle{a-1 } $}
\put (16,36) {$\scriptstyle{a-1 }$}
\put (49,36) {$\scriptstyle{a  } $}
\put(33,-28) {5}}
\multiput(265,0)(53,0){1}{\put (19,-10) {$\scriptstyle{a }$}
\put (40,-10) {$\scriptstyle{a +1 } $}
\put (16,36) {$\scriptstyle{a +1 }$}
\put (49,36) {$\scriptstyle{a  } $}
\put(33,-28) {6}}
 \end{picture}
\vspace{1.5cm}
 \caption{ The six configurations of the six-vertex model, expressed
in terms of spins on the dual lattice.}
 \label{sixvertex}
\end{figure}
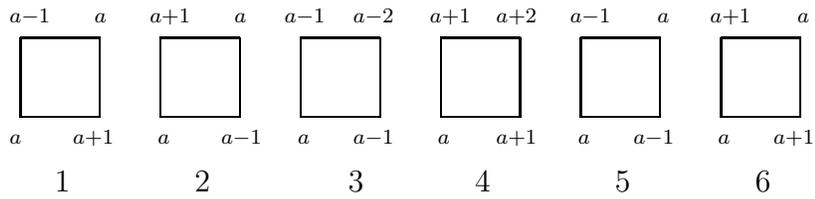

\end{document}